\documentclass[11pt,a4paper]{article}
\usepackage[utf8]{inputenc}
\usepackage{amsmath}
\usepackage{amsfonts}
\usepackage{amssymb}
\usepackage{graphicx}
\usepackage[per-mode=symbol,separate-uncertainty=true]{siunitx}
\usepackage{color}
\usepackage{lineno}
\usepackage{caption}
\usepackage{csquotes}
\usepackage{xspace}
\usepackage{eurosym}
\usepackage{booktabs}
\usepackage{hyperref}
\usepackage{cleveref}
\usepackage{authblk}
\DeclareSIUnit{\sieuro}{\mbox{\euro}}
\xspaceaddexceptions{\grqq \grq \csq@qclose@i \} }
\usepackage[textsize=tiny]{todonotes}

\DeclareSIUnit{\bit}{bit}
\DeclareSIUnit{\samples}{sp}
\DeclareSIUnit{\ppm}{ppm}
\author[1,2*]{Philip Hauer}
\author[2]{Markus Ball}
\author[1,2]{Shania Müller}
\author[2]{Dmitri Schaab}
\author[1,2]{Tim Schüttler}
\author[3]{Rui De Oliveira}
\author[3]{Adam Drozd}
\author[3]{Alexis Rodrigues}
\author[1,2]{Bernhard Ketzer}
\affil[1]{Helmholtz-Institut für Strahlen- und Kernphysik, University of Bonn, Germany}
\affil[2]{Forschungs- und Technologiezentrum Detektorphysik, University of Bonn, Germany}
\affil[3]{CERN, Geneva, Switzerland}
\affil[*]{Corresponding author: hauer@hiskp.uni-bonn.de}

\DeclareSIUnit{\pixel}{px}
\title{GEM Production at the FTD in Bonn}

\usepackage[style=numeric,backend=biber,sorting=none]{biblatex} 
\begin{document}
\maketitle
\begin{abstract}
This manuscript describes the establishment of a local production line for $\SI{10}{\centi\meter} \times \SI{10}{\centi\meter}$ Gas Electron Multiplier (GEM) foils at the Forschungs- und Technologiezentrum Detektorphysik (FTD) at the university of Bonn. 
GEM foils are widely used in modern gaseous detectors, providing high-gain signal amplification and high-rate capability. 
Our fabrication process utilizes a double-mask photolithographic technique on \SI{50}{\micro\meter} polyimide cladded on both sides with \SI{5}{\micro\meter} copper. 
The chemical etching procedure that is required to achieve uniform hole geometries with 
outer hole diameters of \SI{70}{\micro\meter} and inner hole diameters of \SI{50}{\micro\meter} will be described. 

Quality control protocols, including semi-automated optical inspection and high-voltage leakage current tests, demonstrate that foils produced at our facility achieve uniform hole size distributions and leakage currents of less than \SI{1}{\nano\ampere} at \SI{600}{\volt} in air with no discharge hotspots. 
These results confirm that the production chain is capable of delivering high-performance foils suitable for research and development purposes.
\end{abstract}

\section{Introduction}\label{Sec:Introduction}
Since their invention by Fabio Sauli in 1997 \cite{Sauli:1997qp}, Gas Electron Multipliers~(GEMs) have revolutionized the field of Micro-Pattern Gaseous Detectors~(MPGDs). 
By utilizing a \SI{50}{\micro\meter} thin polyimide foil, cladded on both sides  with \SI{5}{\micro\meter} of copper and patterned with a high density of microscopic holes, GEMs allow for electron multiplication within a gas volume under high electric fields.
Their ability to handle particle fluxes exceeding \SI{1e6}{\hertz\per\square\centi\meter} while maintaining excellent spatial resolution and suppressing ion backflow has made them the technology of choice for experiments such as ALICE \cite{ALICETPC:2020ann} and CMS \cite{CMS:2023gfb}.
So far, most of the GEM foils (and other micro-patterned gaseous detectors) have been produced at the CERN Micro-Pattern Technologies (MPT) workshop \cite{Altunbas:2002ds,Villa:2010wj,Abbaneo:2013yua,Giomataris:2004aa}.

As the demand for GEM-based gaseous detectors grows for next-ge\-ne\-ration experiments -- such as the Electron-Ion Collider (EIC) or the High-Luminosity LHC (HL-LHC) -- there is an increasing need for specialized, local production facilities.
Establishing a fabrication chain outside of CERN allows for a better research and development flexibility as rapid prototyping of non-standard foil geometries and materials becomes possible.

The fabrication techniques employed at our facility are based on the standardized production protocols developed by the CERN MPT department.
Through a close and ongoing cooperation with the CERN MPT group, we have successfully implemented their \enquote{recipe} for chemical etching, ensuring that our locally produced foils maintain the same quality and performance standards required for the needs of research and development of future GEM-based particle detectors.
This paper outlines the technical workflow adopted at our institute, from the initial raw material preparation to the final characterization. 
We specifically focus on the chemical etching process to ensure hole uniformity and the implementation of a cleaning protocol essential for stable long-term operation under high voltage.
Furthermore, we have implemented a rigorous, in-house Quality Assurance (QA) chain to ensure that each GEM foil has a suitable quality.
Finally, we present the QA measurements performed during a small-scale production of nine foils.

\section{The Production Process}\label{Sec:ProductionProcess}
\begin{figure}
    \centering
    \includegraphics[width=0.75\linewidth]{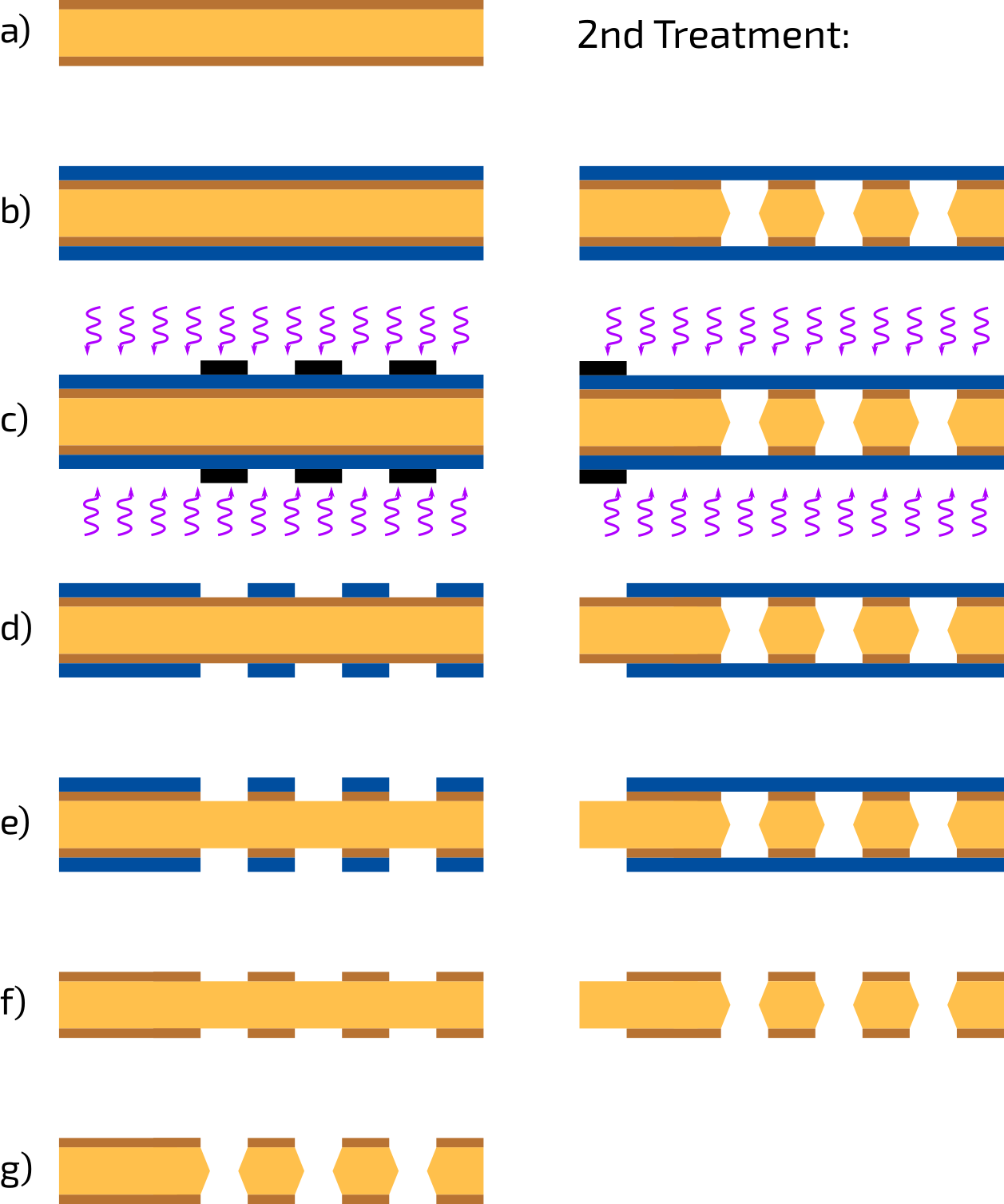}
    \caption{Production process for GEM foils.}
    \label{fig:production_Process}
\end{figure}

The production of GEM foils follows a precise sequence of micro-fabrication steps based on the double-mask technique.
This process is designed to ensure a symmetric biconical structure of the polyimide holes.
A schematic overview of the process is depicted in~\cref{fig:production_Process}.
Details about exact mixtures and concentrations are omitted as the recipe was developed at CERN MPT.

The process begins with a raw composite material consisting of a \SI{50}{\micro\meter} polyimide foil (typically Apical) cladded on both sides with \SI{5}{\micro\meter} of copper~(a).
In order to enhance the adhesion between copper and polyimide, a thin (approximately \SI{10}{\nano\meter} thick) layer of chromium is used between them (not visible in the schematic).
The raw foil is first chemically degreased with acetone to ensure maximum adhesion of the photoresist (a light-sensitive polymer), which is applied to both sides of the foil (b). 
To define the hole matrix, the laminated foil is placed between two high-precision masks. 
These masks contain the negative image of the GEM pattern.
For standard GEM foils, the mask is made from a hexagonal array of circles with a \SI{140}{\micro\meter} pitch.
Next, the \enquote{sandwich} (mask -- photoresist -- copper -- polyimide -- copper -- photoresist -- mask) is exposed to UV light~(c). 
The light hardens the photoresist where it hits, while the areas shaded by the mask remain soft.
Afterwards, the foil is passed through a developer solution (typically sodium carbonate) which washes away the unexposed resist, leaving the copper exposed only where the holes will be etched (d).

With the pattern now physically "printed" onto the foil, the workpiece enters the chemical etching.
First, the copper is etched away (e).
Next, the remaining photoresist is washed away (f).
In order to create the holes in the polyimide, one has to first etch the chromium layer and afterwards through the polyimide (g).
In a last step, the GEM foil should be cleaned thoroughly to remove chemical residues and to stop all reactions.
The whole production chain has to be repeated once more (except for the etching of the polyimide) in order to create the pattern for the high voltage connections (which is denoted with \enquote{2nd treatment} in the sketch).

\section{The Production Facility: FTD at the University of Bonn}\label{Sec:Facility}
The fabrication and characterization of GEM foils are conducted at the Forschungs- und Technologiezentrum Detektorphysik (FTD) of the University of Bonn\footnote{\url{https://www.ftd.uni-bonn.de}}.
The FTD serves as a central hub for detector development not only gaseous, but also semiconductor detectors and calorimeters.
The GEM production as well as the quality assurance takes place in the clean rooms of the FTD, the first performance test in a detector are then conducted in special laboratories.

\subsection{Clean Room Infrastructure}
\begin{figure}
    \centering
    \includegraphics[width=100mm]{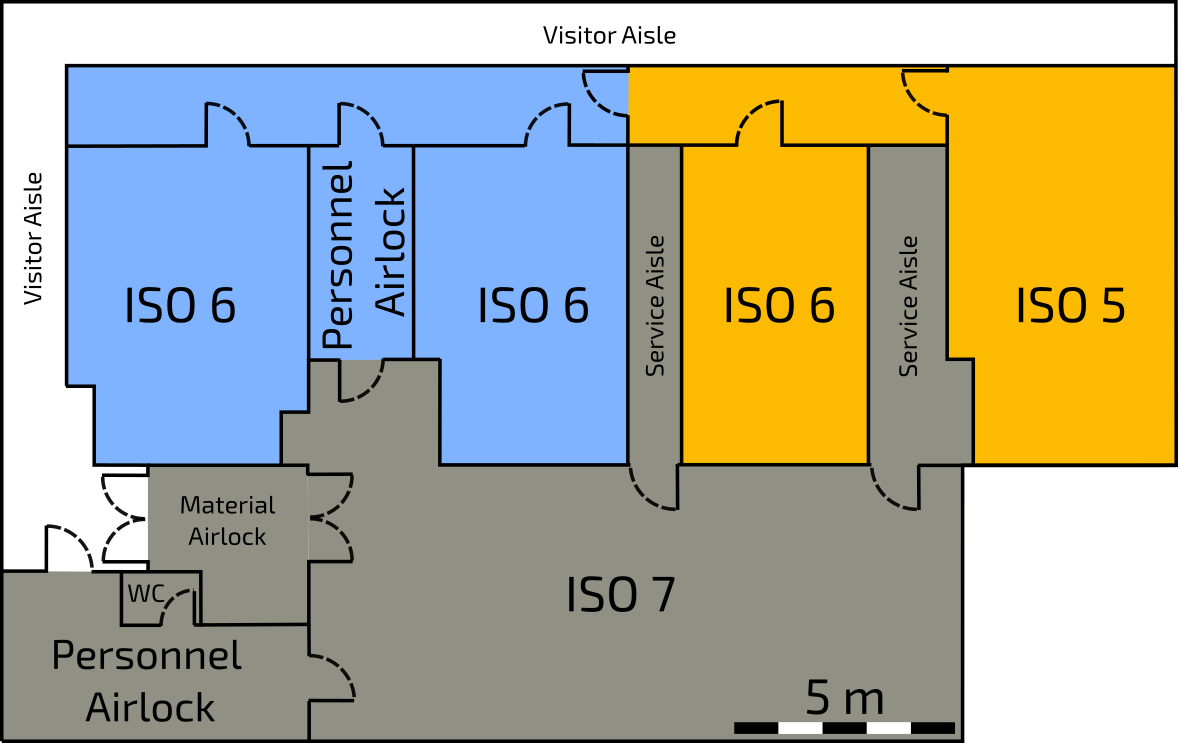}
    \caption{A schematic view of the clean room infrastructure at the FTD.
    All rooms indicated with a yellow color are also only illuminated by special yellow lights in order to perform light sensitive processes (especially photo-lithographic processes) in there.}
    \label{fig:facility:cleanroom}
\end{figure}
The core of the production process takes place within dedicated ISO-certified clean rooms.
Given that dust particles as small as a few micrometers can cause catastrophic electrical breakdowns (shorts) in a GEM foil, the FTD clean rooms are strictly controlled for dust particles, humidity, and temperature.
A schematic view of the clean room infrastructure is depicted in~\cref{fig:facility:cleanroom}.
The light-sensitive steps are all performed under yellow light conditions, depicted in~\cref{fig:facility:cleanroom} with a yellow shade.

\subsection{Processing Laboratories}
Part of the clean rooms are specialized wet-chemistry laboratories equipped specifically for the etching process presented in~\cref{Sec:ProductionProcess}.
In this section, we discuss each production step and also explain where they are performed.

\subsubsection{Application and patterning of photoresist}
In the ISO 6 room in the yellow-light-area is the laminator RLM419p by the company Bungard.
With it, the application of dry film photoresist from both sides of the base material is possible.
We are using a \SI{15}{\micro\meter} thin photoresist Ordyl FP 415.

In order to pattern the photoresist, the Exp 3040 LED by the company Bungard is used.
It exposes the substrate with UV light where the intensity and duration can be set.
As we are using the double-mask technique, we are exposing the material through two masks -- one is below and one above the substrate -- which have to be well aligned (with a precision of $\leq \SI{5}{\micro\meter}$).
The ISO 5 lab is equipped with several wet benches in which the development of the photoresist takes place.
For the development, we are using a Na$_2$CO$_3$-based aqueous solution.

\begin{figure}
    \centering
    \includegraphics[width=0.75\linewidth]{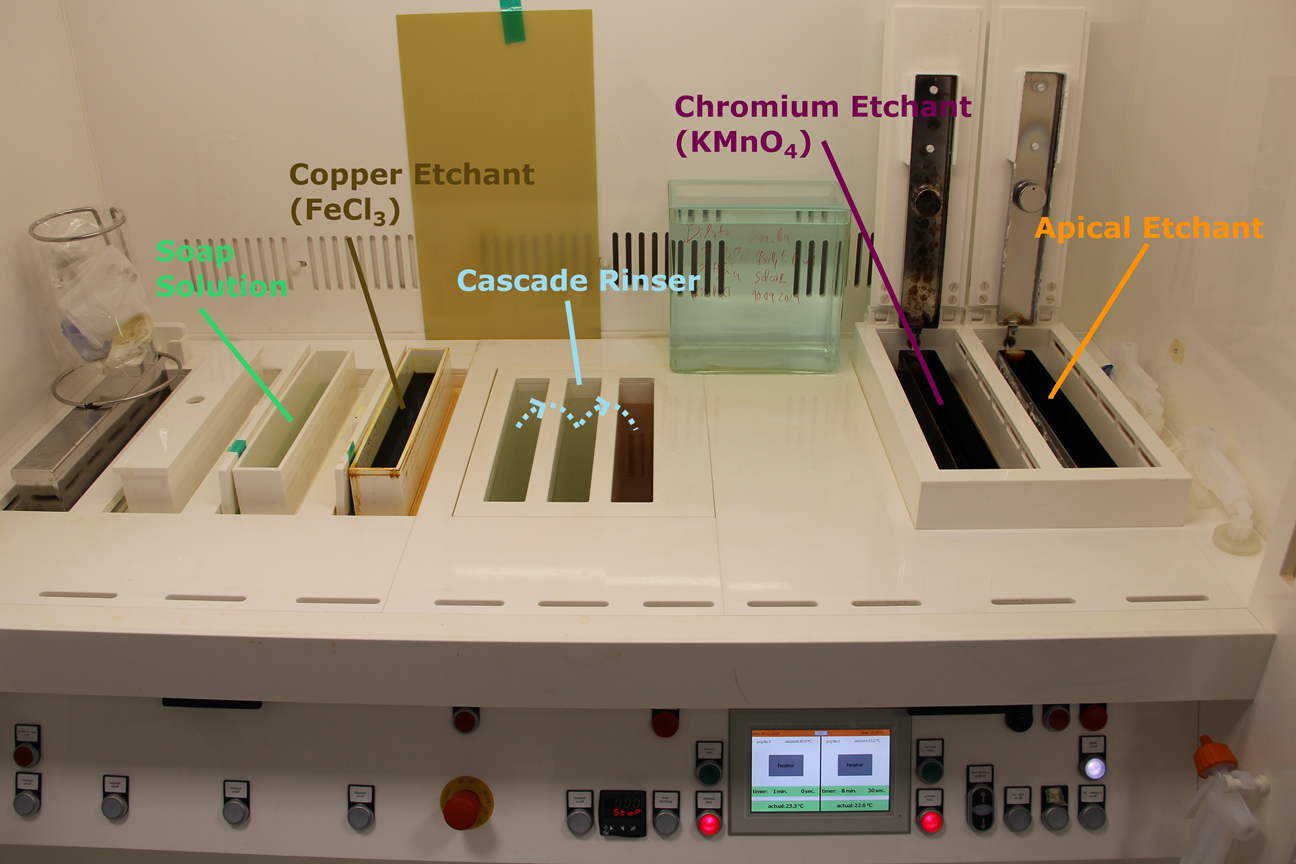}
    \caption{The wet bench for the etching steps.}
    \label{fig:ketzerbank}
\end{figure}

\subsubsection{Etching processes}
The transition to a functional GEM foil requires a multi-stage chemical etching process. 
Following the photolithographic exposure and development of the photoresist, the foils undergo a series of specialized baths at the FTD laboratories to create the characteristic hole pattern as described in~\cref{Sec:ProductionProcess}.
The etching steps are all performed at a dedicated wet bench shown in~\cref{fig:ketzerbank} in the ISO 7 (normal light conditions) as from now on, the sample is not light-sensitive anymore.
The wet-bench is equipped with a cascaded overflow rinser where clean DI water flows into the left beaker, from which it flows into the middle beaker once the left beaker is full and so it continues.
In between each etching process, the foil is placed in the rinser and is also thoroughly sprayed with DI water to ensure that no chemicals are on the sample anymore.

Before primary etching begins, the foils are treated with a dilute soap solution. 
This is an important preparation step, because the openings in the developed photoresist are extremely small (typically around \SI{50}{\micro\meter} wide and \SI{15}{\micro\meter} deep), the high surface tension of standard aqueous solutions would normally prevent the etchants from entering the holes. 
The soap reduces the surface tension and ensures that the subsequent chemical agents can achieve uniform contact with the metal layers.

The first structural etching step removes the copper cladding to define the GEM hole pattern. 
This is performed using a concentrated ferric chloride (FeCl$_3$) solution. 
The process is carefully timed to ensure that the copper is etched completely through to the underlying layers without excessive \enquote{overetching} beneath the photoresist, which would compromise the geometry of the holes.

Once the copper pattern is established, the protective dry-film photoresist is no longer required. 
It is stripped using a mixture of acetone and ethanol. 
This solvent efficiently detaches the photoresist without affecting the copper or the polyimide, leaving a clean, patterned copper surface ready for the next phase.

GEM foils often utilize a thin chromium interface layer between the copper and the polyimide to enhance adhesion. 
This layer must be removed within the hole areas to expose the polyimide for the final etch. 
We use a potassium permanganate (KMnO$_4$) based solution heated to \SI{45}{\degreeCelsius} for this purpose. 

The final and most sensitive step is the creation of the holes in the polyimide substrate. 
To achieve the desired biconical profile we use a chemical mixture consisting of ethylenediamine (EDA), ethanol, potassium hydroxide (KOH) and deionized water.
The KOH and ethylenediamine work in tandem to chemically break down the polyimide chains.
This step is temperature controlled, as it has to be performed at around \SI{65}{\degreeCelsius}.
The duration of this bath and the temperature of the solution are the primary variables used to control the inner diameter of the GEM holes. 

\subsubsection{Final cleaning of the GEM foils}
The first stage involves an immersion in a diluted piranha solution, a potent oxidizing mixture of sulfuric acid (H$_2$SO$_4$) and hydrogen peroxide (H$_2$O$_2$). 
This solution is highly effective at removing organic residues.
Furthermore, it passivates the surface by cleaning the copper surface of oxides and ensures the polyimide walls of the holes are chemically inert.

Following the chemical treatment, the foils are subjected to a high-pressure deionized (DI) water jet. 
Due to the high aspect ratio of the GEM holes, standard immersion rinsing is often insufficient to remove trapped chemical salts or surfactants. 
The high-pressure jet forces DI water through the \SI{50}{\micro\meter} apertures, ensuring a thorough purge of the biconical volume.
In addition, the kinetic energy of the water jet dislodges any microscopic metallic remnants or dust particles that could otherwise deteriorate the high-voltage stability during detector operation.
After the high-pressure rinse, the foils are dried and stored in a dedicated dry cabinet to prevent water spotting.  
The storage is located in the ISO 6 cleanroom on the right side of the second personnel airlock (in \cref{fig:facility:cleanroom}).

\subsection{Quality Assurance}\label{Sec:QA}
To ensure that every foil produced at the FTD meets the rigorous standards required for stable detector operation, we perform a Quality Assurance (QA) protocol. 
This involves a mapping of the geometric properties followed by a high-voltage stress test to verify the electrical integrity.

\subsubsection{Optical QA}\label{Sec:QA:Optical}
The first phase of the QA is an exhaustive optical characterization of the etched hole matrix. 
Given that a standard GEM foil contains approximately \num{6000} holes per \si{\square\centi\meter}, a manual inspection of every hole is unfeasible.
Nevertheless, the first step is always to place the GEM foil on a light table where one can spot peculiarities (e.g. closed holes or defects) by eye.

Next, we utilize a high-precision Keyence digital microscope equipped with image-processing software. 
This system can identify and measure the inner (polyimide) and outer (copper) diameters of every hole within a single frame.
By scanning a few spots of the GEM foil, we generate a statistical distribution of the hole sizes.
Uniformity in these dimensions is paramount, as variations in the hole diameter directly correlate to local variations in the detector gain.
Typically, we are analyzing the hole diameters on five spots -- one in each corner and one approximately in the middle.
On each spot, approximately \num{100} holes can be analyzed.

\subsubsection{Electrical QA}\label{Sec:QA:Electrical}
Once a foil passes the optical inspection, it must prove its stability under high electric fields. 
The electrical QA is performed after the GEM foil has been stored for at least \SI{24}{\hour} in a dry cabinet (relative humidity of $\leq$ \SI{5}{\percent}) to prevent surface moisture from enhancing leakage currents.

For the leakage current test, we apply a high potential difference of \SI{600}{\volt} between the two electrodes of the GEM foil.
An important thing to mention is that we apply the high voltage immediately, i.e. we are ramping the cable (without connection to the GEM) to \SI{600}{\volt} and only after it reached the desired potential, we connect it to the foil.
Concurrently, the leakage current~$I_\textrm{leak}$ is monitored with picoampere precision with a picoammeter presented in~\cite{Utrobicic:2015dxa}.
A high-quality foil typically exhibits a stable leakage current of $\lesssim \SI{1}{\nano\ampere}$. 
A current higher than this, or one that fluctuates wildly, indicates the presence of chemical residues or microscopic dust that must be addressed by further cleaning.

\begin{figure}
    \centering
    \includegraphics[width=0.9\linewidth]{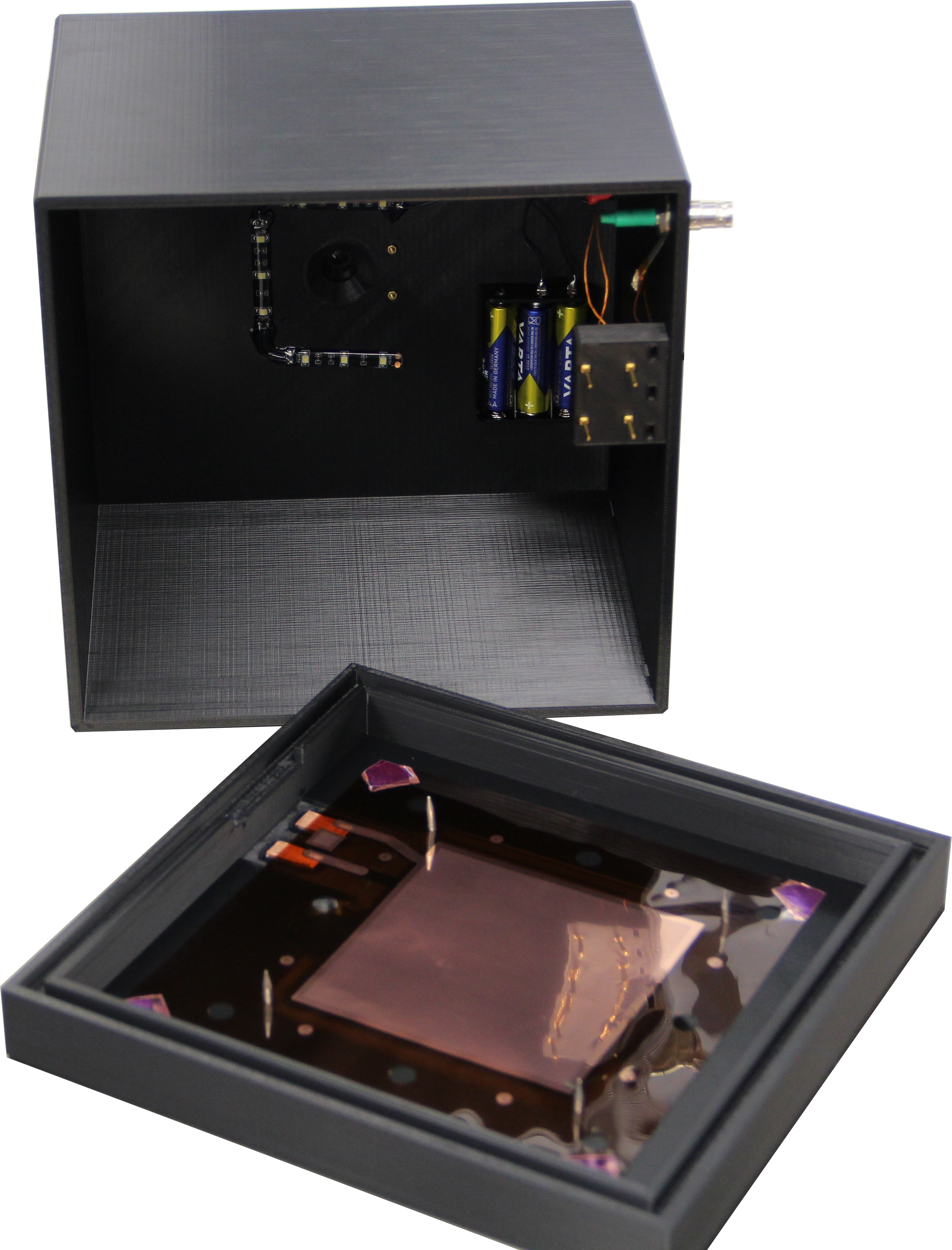}
    \caption{The Spark Detection System (SDS) with an unframed GEM foil.}
    \label{fig:sds}
\end{figure}

During the leakage current measurement, we are utilizing a custom-made spark detection system (SDS) \cite{Ball:2020cbu} as depicted in~\cref{fig:sds}.
It consists of a 3D printed box which can be opened and closed by removing the cover (upper part in the picture) from the base plate.
An SHV connector and feed-through is used to apply high voltage to the GEM electrodes.
The high voltage is guided to spring-loaded pins that are touching the electrodes of the GEM when the SDS is closed.
At the top of the SDS, a commercial webcam is used to record the GEM during the measurement.
With it, we can map the positions of electric discharges on the foil.
If too many sparks are occuring on one spot, we also reclean or discard the GEM foil \cite{Merlin:2016qac}.
The LED strips at the top are used to illuminate the box from the inside to crosscheck the position of the GEM once the system is closed.
For the measurement, the LEDs are turned off again.

\subsubsection{Yield}\label{Sec:QA:Yield}
While the primary goal of the GEM production at the FTD is research and development, including the prototyping of novel geometries and the testing of alternative substrate materials, establishing a reliable baseline for production of standard GEM foils, the yield is essential.
A high yield indicates not only the quality of the infrastructure but also the maturity of the technology transfer from CERN.
To validate the stability of our workflow, we conducted a dedicated production focusing on a series of nine standard GEM foils.
It should be noted, however, that nine foils constitute a limited sample size, and a larger production run would be needed to draw more robust statistical conclusions about the long-term consistency of the process.

For this specific run, all fabrication parameters were kept constant. 
This allowed us to move beyond individual \enquote{best-case} results and evaluate the statistical reliability of the FTD production line.
For each foil, the quality assurance protocol has been conducted as explained in~\cref{Sec:QA}.

\begin{figure}
    \centering
    \begin{minipage}{0.45\textwidth}
    \includegraphics[width=\linewidth]{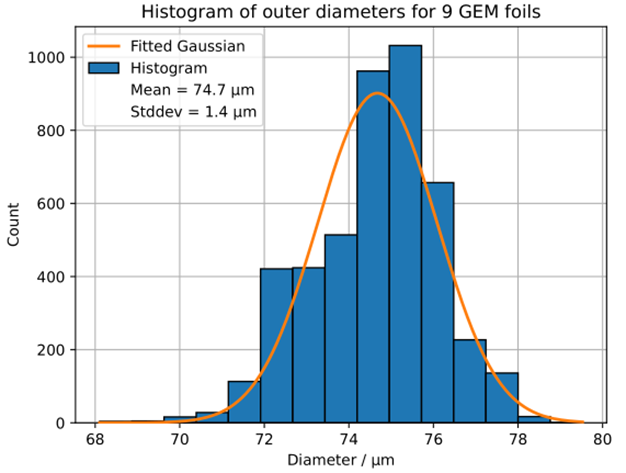}
    \caption{Outer hole size distribution.}
    \label{fig:QA:OuterHoles}
    \end{minipage}\hfill
    \begin{minipage}{0.45\textwidth}
    \centering
    \includegraphics[width=\linewidth]{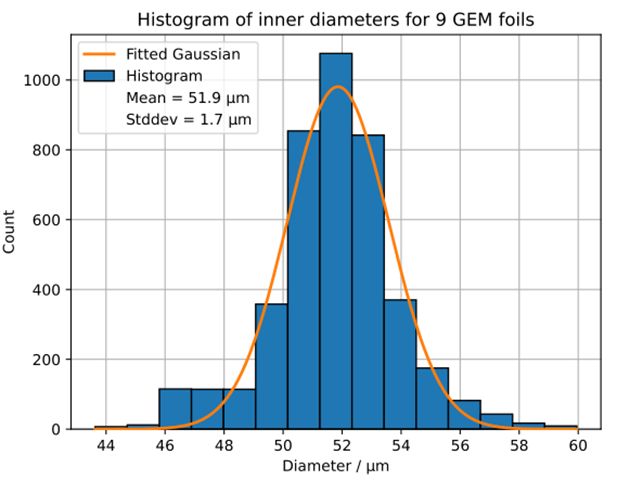}
    \caption{Inner hole size distribution.}
    \label{fig:QA:InnerHoles}
    \end{minipage}
\end{figure}

When scanning the foils by eye for peculiarities, we observed that one foil (number 05) showed a big discolored spot.
All other foils looked very good and we proceeded with all nine foils to the inspection by microscope.
The results of the optical QA of these nine foils are shown in \cref{fig:QA:InnerHoles,fig:QA:OuterHoles}.
The outer holes are on average \SI{74.7 \pm 1.4}{\micro\meter} in diameter, while the inner holes have a diameter of \SI{51.9 \pm 1.7}{\micro\meter}\footnote{The uncertainty given here is the standard deviation of all measurements.}.
These values are averaged over all nine foils while for each foil approximately \num{500} holes were analyzed.
As the design values for the outer and inner diameter are \SI{70}{\micro\meter} and \SI{50}{\micro\meter}, respectively, we have a slight over-etching for the outer diameter, while we are (taken the statistical uncertainties into account) within the design value of the inner diameter. 

\begin{figure}
    \centering
    \begin{minipage}[b]{0.45\textwidth}
    \includegraphics[width=\linewidth]{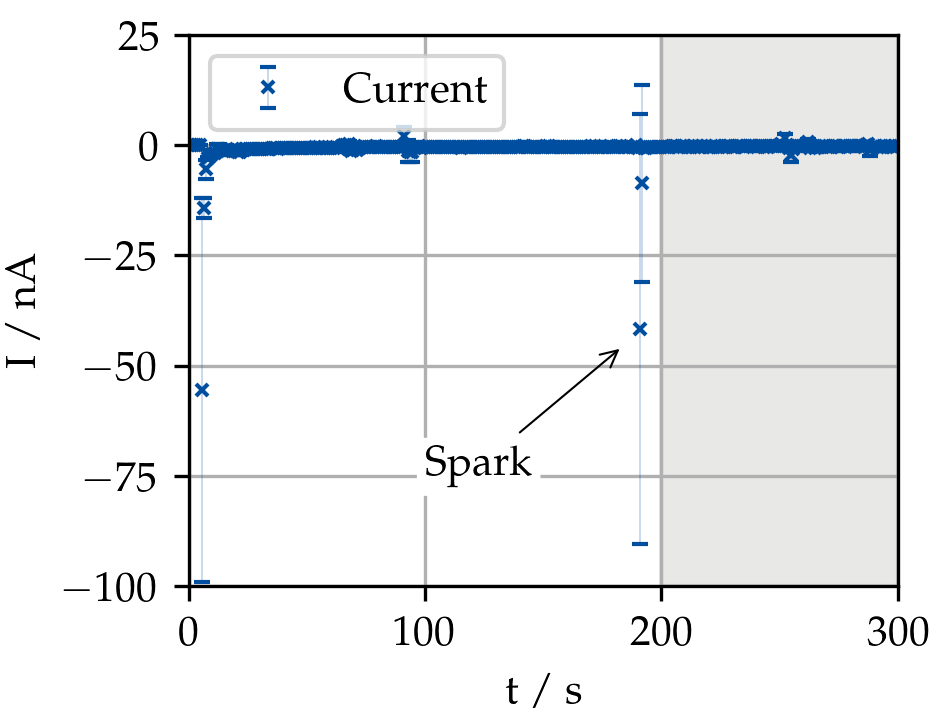}
    \caption{Leakage current measurement of foil 07 at \SI{650}{\volt}.}
    \label{fig:QA:leakCurrent:foil07}
    \end{minipage}\hfill
    \begin{minipage}[b]{0.45\textwidth}
    \includegraphics[width=\linewidth]{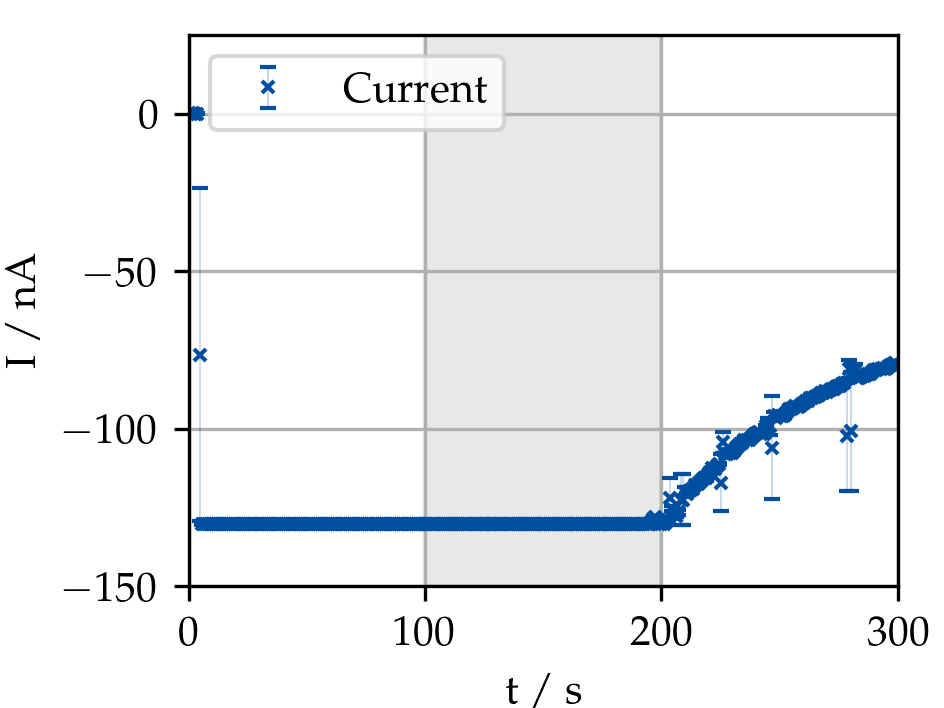}
    \caption{Leakage current measurement of foil 05 at \SI{600}{\volt}.}
    \label{fig:QA:leakCurrent:foil05}
    \end{minipage}
\end{figure}

The next step in the quality assurance protocol is the leakage current measurement.
The results for each foil is shown in~\cref{tab:yield:electricQA}.
While most foils pass the requirement of the electric QA (as exemplarily shown in \cref{fig:QA:leakCurrent:foil07}), foil 05 clearly did not pass this step as the measured leakage current was \SI{130}{\nano\ampere} (depicted in \cref{fig:QA:leakCurrent:foil05}), which is the maximum value that our picoammeter can measure.
Because of this, we stopped the measurement after \SI{200}{\second} which explains the drop of the current afterwards.
The leakage current given in \cref{tab:yield:electricQA} is always an average over the last \SI{100}{\second} of a measurement (as indicated by the grey area in \cref{fig:QA:leakCurrent:foil05,fig:QA:leakCurrent:foil07}).

\begin{figure}
    \centering
    \begin{minipage}[b]{0.45\textwidth}
    \centering
    \includegraphics[width=\linewidth]{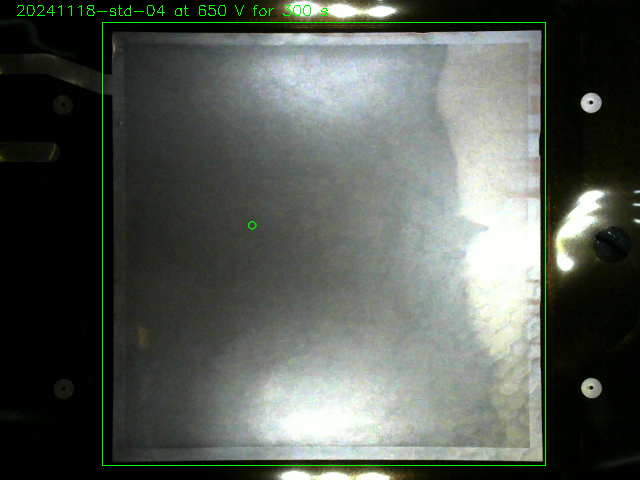}
    \caption{Spark map of foil 07, with only one spark in total.}
    \label{fig:QA:sparkMap:foil07}
    \end{minipage}\hfill
    \begin{minipage}[b]{0.45\textwidth}
    \centering
    \includegraphics[width=\linewidth]{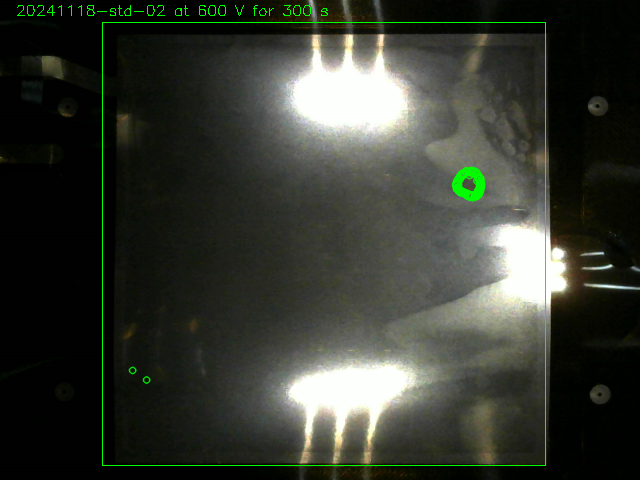}
    \caption{Spark map of foil 05, with 87 spark in total.}
    \label{fig:QA:sparkMap:foil05}
    \end{minipage}
\end{figure}

Consecutively, the SDS was running and recorded all sparks.
The output of a good foil is exemplarily depicted in \cref{fig:QA:sparkMap:foil07}.
Again, for all foils, the number of sparks and also the spatial distribution was inconspicuous (i.e. if there were many sparks, they were all distributed across the whole GEM foil) except for foil~05.
The sparks were mostly around the same spot (as shown in \cref{fig:QA:sparkMap:foil05}) that already caught our attention during the coarse optical scan with the light table.
We therefore discarded foil 05 and we end up with eight working GEM foils.

\begin{table}
    \sisetup{parse-numbers = false}
    \centering
    \caption{Results of the electric QA.}
    \begin{tabular}{S[table-format=2.0]
    S[table-format=5.3]@{\,\( \pm \)\,}
    S[table-format=1.3]
    S[table-format=2.0]
    S[table-format=3.0]}
    \toprule
        {GEM foil} & \multicolumn{2}{c}{$I_\textrm{leak}$ / nA} & {Nr. of sparks} & {Voltage $U$ / V} \\ \midrule
         01 & -1.0    & 0.2   &  3 & 550 \\
         02 & -0.756  & 0.002 &  1 & 600 \\
         03 & -0.395  & 0.005 & 24 & 600 \\
         04 & -0.419  & 0.003 &  0 & 600 \\
         05 & -130.36 & 0.02  & 87 & 600 \\
         06 & -0.286  & 0.002 &  3 & 600 \\
         07 & -0.32   & 0.02  &  1 & 650 \\
         08 & -0.531  & 0.002 &  3 & 600 \\
         09 & -0.121  & 0.003 &  0 & 600 \\ \bottomrule
    \end{tabular}
    \label{tab:yield:electricQA}
\end{table}


\section{Summary and Outlook}\label{Sec:SummaryOutlook}
In this work, we have detailed the successful establishment of a high-precision production and characterization chain for Gas Electron Multiplier (GEM) foils at the Forschungs- und Technologiezentrum Detektorphysik (FTD) of the University of Bonn. 
By implementing the \enquote{CERN recipe} for chemical etching and cleaning, and adapting it to our local infrastructure, we have demonstrated that GEM foils can be fabricated with our university laboratory setting.

The results presented in this paper highlight several key milestones.
While we already produced many GEM foils for our own needs, we confirmed with a production run of nine foils that the FTD facility can maintain high consistency, with leakage currents typically remaining below \SI{1}{\nano\ampere} at \SI{600}{\volt}, making these foils fully compatible with the requirements of major particle physics experiments.
Currently, we are limited to the size of $\SI{10}{\centi\meter} \times \SI{10}{\centi\meter}$ GEM foils.
Nevertheless, they can still be used for R\&D purposes.

With the production baseline now firmly established, the FTD is well-positioned for the next phase of gaseous detector development. 
Future activities will focus on investigating alternative metal cladding compositions to minimize material budget for low-energy physics applications.
Furthermore, we want to work on the optimization of our quality assurance procedure by implementing a large-scale XY-microscope with which one can scan each hole of a GEM foil and potentially even spot defects, similar to \cite{Hilden:2015qpi}.

A significant advantage of establishing a GEM production facility within a university environment is the opportunity for specialized student training in advanced photolithography. 
This educational synergy was successfully demonstrated during the DRD1 School 2025\footnote{https://indico.cern.ch/event/1522987/} and is further utilized through dedicated hands-on seminars integrated into the University of Bonn’s official physics curriculum.
    
In conclusion, the close cooperation with CERN MPT and the specialized infrastructure at the University of Bonn have created a robust environment for GEM technology. 
The FTD can now act as a contributor to the global micro-pattern gaseous detector community, capable of both cutting-edge R\&D and reliable small-scale production.


\printbibliography
  
\end{document}